\documentclass[aps,prc,reprint,superscriptaddress,showpacs,amsmath,amssymb,longbibliography,lengthcheck]{revtex4-1}

\usepackage[dvips]{graphicx}
\usepackage{mathrsfs}
\usepackage{txfonts}

\begin{document}

 \title{Cluster formations in deformed states for $^{28}$Si and
 $^{32}$S} 

\author{Takatoshi Ichikawa}%
\affiliation{Yukawa Institute for Theoretical Physics, Kyoto University,
Kyoto 606-8502, Japan} 
\author{Yoshiko Kanada-En'yo}
\affiliation{Department of Physics, Kyoto
University, Kyoto 606-8502, Japan} 
\author{Peter M\"oller}
\affiliation{Theoretical Division, Los Alamos National Laboratory, Los
Alamos, New Mexico 87545, USA} 
\date{\today}

\begin{abstract}
 We study cluster formation in strongly deformed states for $^{28}$Si
 and $^{32}$S using a macroscopic-microscopic model.  The study is based
 on calculated total-energy surfaces, which are the sums of
 deformation-dependent macroscopic-microscopic potential-energy surfaces
 and rotational-energy contributions. We analyze the
 angular-momentum-dependent total-energy surfaces and identify the
 normal- and super-deformed states in $^{28}$Si and $^{32}$S,
 respectively. We show that at sufficiently high angular momenta
 strongly deformed minima appear. The corresponding microscopic density
 distributions show cluster structure that closely resemble the
 $^{16}$O+$^{12}$C and $^{16}$O+$^{16}$O configurations. At still higher
 deformations, beyond the minima, valleys develop in the calculated
 surfaces. These valleys lead to mass divisions that correspond to the
 target-projectile configurations for which molecular resonance states
 have been observed. We discuss the relation between the one-body
 deformed minima and the two-body molecular-resonance states.
\end{abstract}


\maketitle

A rich variety of nuclear structure data in the $s$-$d$ shell region
provides an excellent opportunity to investigate how a system
transitions between one-body-like mean-field and two-body-like cluster
structures~\cite{Bet97,Fre07}.  Because of recent progress in
experimental techniques, it has been possible to determine that strongly
deformed states exist in $^{36}$Ar and $^{40}$Ca, by the observation of
$\gamma$-ray cascades typical of rotational bands~\cite{Sve00,Ide01}.
These bands are called super-deformed (SD) bands. Such new have
triggered renewed interest in whether the strongly deformed states exist
in other $s$-$d$ shell nuclei.

In this connection, the existence of such states in $^{28}$Si and
$^{32}$S has been theoretically
suggested~\cite{Matsu,Ina03,Kim04,Tani09}.  Many experimental
searches for, and studies of such states have been
performed~\cite{Pan10,Lon10}.  An important feature of nuclear structure
in the $s$-$d$ shell region is that the densities of strongly-deformed
one-body states often exhibit significant cluster
structure~\cite{Hori10,enyo10}, similar to the $^{16}$O+$^{16}$O
configuration suggested to exist in $^{32}$S~\cite{Mori85,Kim04,Ohk02}.
Recently, the existence of  alpha-cluster states in $^{32}$S was
clearly shown in elastic $^{28}$Si+$\alpha$ scattering experiments~\cite{Lon10}.
However, the existence of the strongly deformed states and the mechanism
of the cluster formations in $^{28}$Si and $^{32}$S have not yet been well
established.

Another important observation is the molecular resonances emerging just
below the Coulomb barrier in the two-body entrance channel, in both the
$^{16}$O+$^{12}$C and $^{16}$O+$^{16}$O reactions leading to $^{28}$Si
and $^{32}$S, respectively.  That is, the molecular-resonance states
would consist of the $^{16}$O+$^{12}$C and $^{16}$O+$^{16}$O cluster
components, similar to the clusters in the strongly deformed states in
$^{28}$Si and $^{32}$S.  It is thus interesting to investigate the
relation between the one-body deformed states and the two-body
molecular-resonance states and the association with the cluster
formations in the deformed states.

Two different theoretical approaches have been used to describe the
deformed states in the $s$-$d$ shell nuclei.  One is nuclear structure
calculations using one-body wave functions.  Leander and Larsson
identified several distinct minima with exotic shapes using the
macroscopic-microscopic model~\cite{Lean75}.  Minima at high angular
momenta were also investigated based on a cranking model for the
rotational inertia~\cite{Rag81}.  However, the $\ell^2$ term in their
mean-filed potential leads to many unphysical minima at large
deformations. Moreover, strongly necked-in shapes are not possible in
the Nilsson perturbed-spheroid ($\epsilon$) parametrization.

The SD states and the low-lying excited states have also been treated in
Hartree-Fock-type (HF) self-consistent mean-field
calculations~\cite{Matsu,Kane09}, often coupled with the
generator-coordinate method (GCM)~\cite{Rod00,Ben03}.  For $^{32}$S,
Kimura and Horiuchi suggested the existence of an SD band containing the
$^{16}$O+$^{16}$O cluster components, based on the antisymmetrized
molecular dynamics coupled with GCM~\cite{Kim04}.  They also showed that
a third rotational band with $N=28$, where $N$ is the principal quantum
number of the relative motion between clusters, in the $^{16}$O+$^{16}$O
configuration connects to the molecular-resonance states. The existence
of the normal-deformed (ND) state in $^{28}$Si and its relation to the
$^{16}$C+$^{12}$O molecular resonances were also
investigated~\cite{Enyo04,Tani09}.

The second approach is reaction calculations using a two-body potential
model appropriate to the entrance channel. Those studies are mainly
based on an optical potential which well reproduces the experimental
elastic or inelastic cross
sections~\cite{Sch70,twobody1,twobody2,Kon89,Ohk02,Ohk04}.  For
$^{32}$S, Ohkubo and Yamashita~\cite{Ohk02} calculated the SD bands with
the deep $^{16}$O-$^{16}$O potential~\cite{Ohk02}.  They recognized
three rotational bands with $N=24$, 26, and 28, and showed that the
lowest and the third bands correspond to the SD band and the molecular
resonances, respectively.  Kocak {\it et al.} also obtained a similar SD
band with $N=24$ using the alpha-alpha double-folding
potential~\cite{koc10}.

The aim of this paper is to show cluster formations in the strongly
deformed states for $^{28}$Si and $^{32}$S. We find that consideration
of rotational contributions to the energy is essential.  In the study
here we apply for the first time the macroscopic-microscopic model, so
successful in the description of fusion and fission reactions in
heavy-mass systems~\cite{Nix68,Nix69,Bol72,Mol95,Mol09}, to very light
nuclei.  The model allows us to describe both one-center deformed and
two-center cluster-like configurations with mass asymmetry within the
same model framework.  In this approach the clusters are joined by a
neck region with a lower single-particle density.  For lighter mass
systems, such treatments are essential for a unified description of the
whole process, because the scission point shape 
closely resembles that of the saddle point, as has been well
established in fusion-fission processes below the Businaro-Gallone
point~\cite{san99}.  We calculate and analyze total-energy surfaces
which are the sums of a potential-energy surface and a rotational-energy
contribution, which both are functions of five shape-degrees of freedom.
We use the immersion technique to identify reaction channels that we
expect correspond to molecular resonances including various mass
asymmetric divisions.  We show that in this model minima with density
distributions corresponding to the cluster configurations of
$^{16}$O+$^{12}$C and $^{16}$O+$^{16}$O appear at high angular momenta.

\begin{figure}[t]
\includegraphics[keepaspectratio,width=\linewidth]{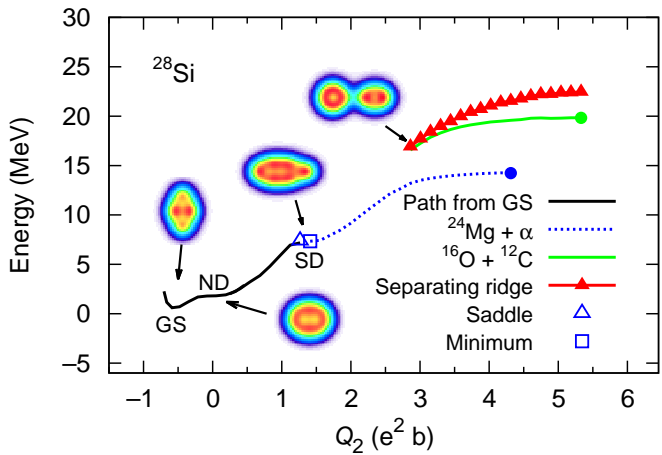}\\%
\caption{(Color online) Potential energy curve for $^{28}$Si versus the
quadrupole moment. The open square and SD denote the super-deformed
minimum.  The absolute minimum is denoted by GS.  The solid line denotes
the potential versus $Q_{\rm 2}$ near the ground state, along a
trajectory that locally minimizes the energy.  The gray (green) and
dotted lines denote the only relatively prominent valleys found in the
one-body potential-energy surface. They correspond to shapes with
asymmetries similar to the $^{16}$O+$^{12}$C and $^{24}$Mg+$\alpha$
reaction channels. The solid line with superimposed triangles is the
ridge separating these two channels.}
\end{figure}
\begin{figure}[t]
\includegraphics[keepaspectratio,width=\linewidth]{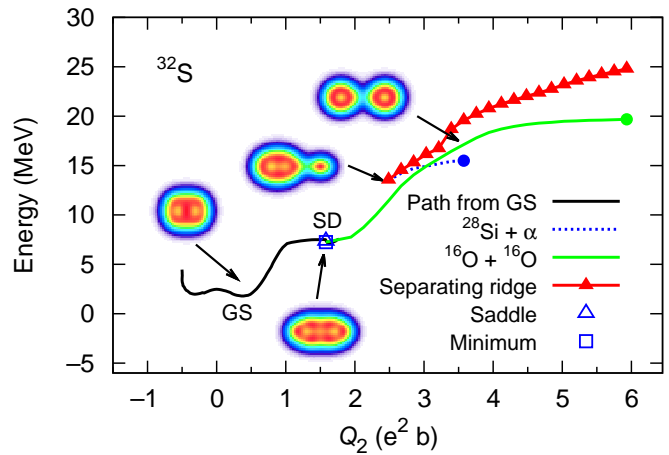}\\%
\caption{(Color online) Potential energy curve for $^{32}$S versus the
quadrupole moment. The gray (green) and dotted lines denote paths leading to
valleys in the one-body potential energy surface corresponding to
strongly necked-in one-body shapes with asymmetries similar to those of
the $^{16}$O+$^{16}$O and $^{28}$Si+$\alpha$ reaction channels.  The
other symbols are the same as Fig. 1.}
\end{figure}

We use the three-quadratic-surface (3QS)
parametrization~\cite{Nix68,Nix69} to describe nuclear shapes in a
five-dimensional deformation space. The shape degrees of freedom are a
quadrupole-moment parameter $Q_2$, a neck-related parameter $\eta$,
left- and right-fragment deformation parameters, $\epsilon_{\rm f1}$ and
$\epsilon_{\rm f2}$, and a mass-asymmetry parameter $\alpha_{\rm g}$.
The parameter $\eta$ describes the curvature of the middle body.  The
parameter $\epsilon$ is the Nilsson perturbed-spheroid parameter.  Near
scission we have to a very good approximation $\alpha_{\rm g}=(M_1 -
M_2)/(M_1 + M_2)$, where $M_1$ and $M_2$ are the masses of the left and
right nascent fragments, respectively.  The microscopic single-particle
potential is calculated by folding a Yukawa function over the shape or
``sharp-surface generating volume'' \cite{Bol72}.

We calculate the adiabatic one-body potential-energy surface in a
five-dimensional deformation space for $^{28}$Si and $^{32}$S and and
analyze their structure using the immersion method.  Details of the
model are given in Ref.~\cite{Mol09}.  The parameters correspond to
FRLDM(2002)~\cite{Mol04}.  We calculate the potential energies at
$41\times15\times15\times15\times35$ grid points for $Q_2$, $\eta$,
$\epsilon_{\rm f1}$, $\epsilon_{\rm f2}$, and $\alpha_{\rm g}$,
respectively.  For $\alpha_{\rm g}$ grid points we use $-0.025 (0.025)
0.825$; the fragment shape grid points are the same as
Ref.~\cite{Mol09}; in $\eta$ the choice is similar.  We take into
account the shape dependence of the $A^{0}$ and Wigner terms in our
calculations~\cite{Mol89}.
However, in the form introduced in our model
the Wigner energy is zero for the $N=Z$ nuclei we consider here.
Near the ground states, we perform
$\beta$-constrained calculations, which describe better one-body shapes
for small deviations from spherical shape.  For the purpose of comparing
with calculations in other shape parameterizations we sometimes give the
deformations of our shapes in terms of the $\beta$ shape parameters,
obtained by expanding the 3QS shapes in spherical harmonics
\cite{Mol95}.  We calculate nuclear density distributions and determine
the number of nucleons in the left and right fragments by integrating
the single-particle densities~\cite{ich09}.

Figures 1 and 2 show the calculated results for $^{28}$Si and $^{32}$S
as ``optimal'' one-dimensional potential-energy curves imbedded in the
five-dimensional space, versus the quadrupole moment.  Nuclear densities
at points of special interest are also given.  The calculated
potential-energy curves for $^{28}$Si and $^{32}$S are quite similar to
other calculations~\cite{Kim04,Tani09,Matsu,Ben03,Rod00,Kane09}.  At
larger $Q_2$ valley-like structures appear in the 5D surface; we show
curves corresponding to the bottom of the only two relatively prominent,
that is deep and persistent, valleys we identify.  The scission points
in each reaction channel are denoted by solid circles. The one-body
ground state connects continuously to these two-body cluster channels.

For $^{28}$Si, we identify two paths: one given by the dotted line,
leading to the $^{24}$Mg+$\alpha$ reaction channel, and a second, given
by the gray line, leading to $^{16}$O+$^{12}$C reaction channel.  Those
are separated by a potential ridge, shown as a solid line with
superimposed triangles.  The calculated ground-state shape is oblate
with $Q_2=-0.59$ ($e^2$b).  We obtain a flat potential energy curve near
$Q_2=0.05$ ($e^2$b), which is consistent with the HF calculation of
Ref.~\cite{Kane09}.  Although this flat area corresponds to a much
smaller beta than the $\beta_2\sim0.5$ of the ND minimum found in
Ref.~\cite{Tani09} we label this flat part at $Q_2=0.05$ ($e^2$b) ND.
For higher angular momenta it evolves into a more well-localized
minimum.  We furthermore identify the additional energy minimum at
$Q_2=1.41$ ($e^2$b) at $\beta=0.68$, denoted by the open square, with
the SD minimum \cite{Tani09}.  In spherical shell-model terminology this
is interpreted as a $4p$-$16h$ ($4\hbar\omega$) state with the intruder
single-particle orbital of $1/2[330]$ (labeled with the Nilsson
asymptotic quantum numbers $\Omega^{\pi}[Nn_z\Lambda]$) at the Fermi
energy for both protons and neutrons.  From our {\it deformed}
mean-field model point of view there are no particle-hole excitations,
since this is the lowest possible energy at this deformation.
In this sense, our calculated results and those at $J^{\pi}=0^+$ of
Ref.~\cite{Tani09} are quite similar to each other, both 
as relates to  the
shape configurations at the ground-state, the ND, and the SD minima and
to the single particle configurations (see.~(a), (b), and (c) of
Fig.~2 and (a) and (b) of Fig.~4 in Ref~\cite{Tani09}).
The  optimal potential-energy curves obtained by \cite{Tani09}
for the $^{24}$Mg+$\alpha$ and $^{16}$O+$^{12}$C channels in $^{28}$Si
are also quite similar to the results here.

\begin{figure}[t]
\includegraphics[keepaspectratio,width=\linewidth]{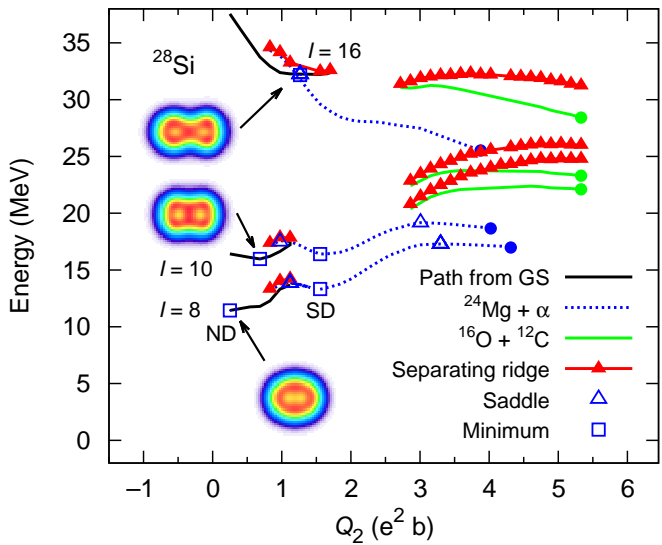}\\%
\caption{(Color online) Potential energy curves versus quadrupole moment
for $I=8$, 10, and 16 for $^{28}$Si .  The shape configuration of the ND
 minimum is changed to the $^{16}$O+$^{12}$C like cluster one at $I=10$.
 The symbols are the same as Fig. 1.}
\end{figure}

\begin{figure}[t]
\includegraphics[keepaspectratio,width=\linewidth]{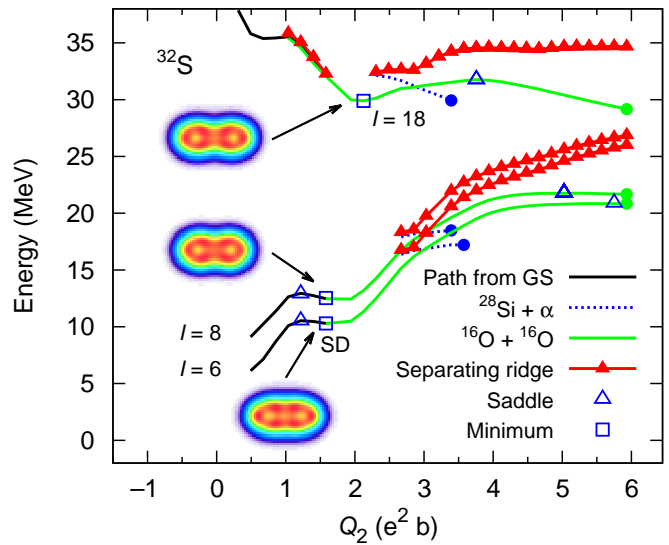}\\%
\caption{(Color online) Potential energy curves at the intrinsic angular
 momentum $I=6$, 8, and 18 for $^{32}$S versus the quadrupole 
 moment.
The shape configuration of the SD minimum is changed to the
 $^{16}$O+$^{16}$O like cluster one at $I=8$.
The symbols are the same as Fig. 2.}
\end{figure}
 
For $^{32}$S, we identify two paths, one leading to the
$^{16}$O+$^{16}$O (the gray line) the reaction channel, the other to the
$^{28}$Si+$\alpha$ (the dotted line) reaction channel, and the
separating ridge (the solid line with the filled triangles).  The
calculated ground-state is prolate with $Q_2=0.39$ ($e^2$b)
corresponding to $\beta_2=0.24$.  We also obtain an additional, almost
symmetric minimum at $Q_2=1.58$ ($e^2$b) corresponding to $\beta=0.72$,
denoted by an open square. This minimum is the SD state.  Again, in
spherical shell-model terminology this is interpreted as a $4p$-$12h$
($4\hbar\omega$) state with the intruder single-particle orbital
$1/2[330]$ at the Fermi energy for both protons and neutrons.

We now calculate the total energy versus angular momentum (and $Q_2$)
for $^{28}$Si and $^{32}$S. We calculate the macroscopic rigid-body
moment of inertia for the shapes of interest and obtain the total energy
by adding the shape-dependent rotational energy to the five-dimensional
potential-energy surface.  The rotational energy $E_{\rm R}$ is then
given by $E_{\rm R}=\hbar^2 I (I+1)/2 \mathscr{J}_\perp$, where $I$
denotes the collective rotational angular momentum in the intrinsic
frame and $\mathscr{J}_\perp$ denotes the rotational moment of
inertia. We only consider rotations around the $\rho$ axis, which is
perpendicular to the symmetry axis ($z$ axis)~\cite{Has88}.  Even if the
two fragments are well separated, we treat such configurations as
rigid-body rotors.  We analyze the total potential-energy surfaces
obtained at each $I$, using the water immersion method.

\begin{figure}[t]
\includegraphics[keepaspectratio,width=\linewidth]{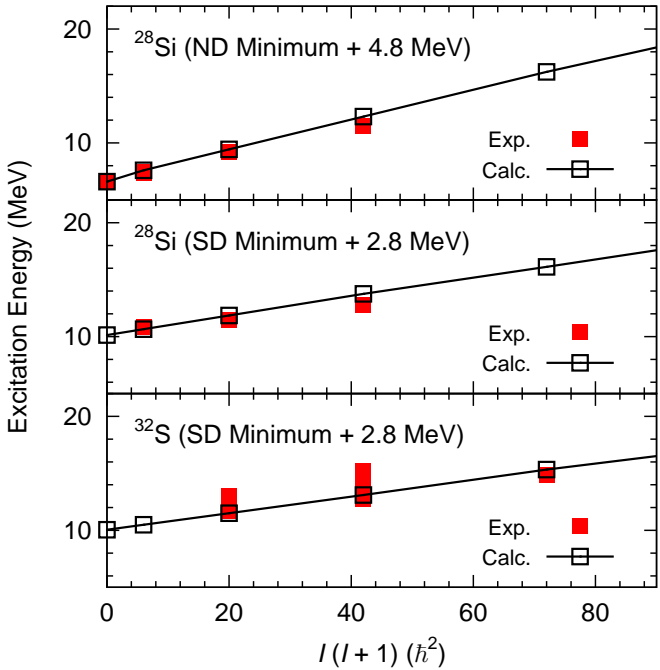}\\%
\caption{(Color online) Rotational levels in the ND and SD minima versus
the intrinsic angular momentum $I$ for $^{28}$Si and $^{32}$S.  The
solid line with the open square denotes the total energies at the ND or
SD minima.  The solid square denotes experimental data taken from
Refs.~\cite{Mori85,She82} We normalize the calculated band-head energies
to experimental data by shifting the ND minimum for $^{28}$Si by $+4.8$
MeV and the SD minima for $^{28}$Si and $^{32}$S by $+2.8$ MeV.  }
\end{figure}

Figures 3 and 4 show, for three different angular momenta, total
 potential energy curves, along one-dimensional ``minimal-energy'' paths
 imbedded in the five-dimensional deformation space for $^{28}$Si and
 $^{32}$S, respectively.  The ND and SD minima are present for $^{28}$Si
 and the SD minimum for $^{32}$S.  In the figures, the ND and SD minima
 at each $I$ are indicated by open squares.  The dotted lines through
 these minima are the optimal pathways from ground-state-like shapes to
 the $^{24}{\rm Mg} + \alpha$ and $^{28}{\rm Si} + \alpha$ channels.
 The other symbols are the same as Figs.~1 and 2. Nuclear densities at
 points of special interest are also given.  The potential pockets at
 the ND minima vanish as the angular momentum increases.

Figure 5 shows the calculated rotational bands in the ND and SD minima.
In the figure, the total energies at the ND and SD minima are denoted by
solid lines with open squares.  For 
 comparison, we plot experimental data
as solid squares.  However, experimental band assignments have not been
well confirmed, except for the ND state of $^{28}$Si and the
alpha-cluster states of $^{32}$S.

For the ND state of $^{28}$Si, the band assignments have been confirmed
by $\gamma$-ray measurements~\cite{Gla81}.  We thus directly compare our
calculated results with those data.  In the top panel of the figure, we
normalize the calculated bandhead energies to the lowest levels of the
experimental data, because our calculated energies show some discrepancy
with respect to the experimental data. After this normalization, the
calculated level spacings agree well with data, which suggests that the
calculated deformation of the ND minimum is realistic. In comparison to
other calculations, we find that our calculated rotational bands of
the ND state for $^{28}$Si correspond to those with the lowest $N$, 
namely $N=18$ in Refs.\ \cite{Enyo04,Ohk04}.

For the alpha-cluster states of $^{32}$S, those were recently clearly
identified in $^{24}$Mg+$\alpha$ elastic 
scattering experiments.  
However, we do not identify  minima  corresponding to
those states in our calculations.  The experimentally deduced moment of
inertia for those states is about two times as large as our calculated
results of the SD states (see the bottom panel of Fig. 5), indicating
that it is necessary to take into account rotations at smaller
$Q_2$ than in the present calculations, in order to
reproduce this experimental result. At such small $Q_2$, triaxial
deformations are important. To access triaxial shapes a model extension 
such as Ref.\ \cite{Mol09,Mol11} is necessary. After such extension,
the Jacobi shape transitions in the $\beta$-$\gamma$ deformation space,
as shown in Ref.\ \cite{Pan10}, could be studied. 

The rotational bands for the SD states of $^{28}$Si and
$^{32}$S have been not confirmed yet. Therefore we are limited
to plotting possible candidates
proposed by Refs.\ \cite{Mori85,She82} for those states.  The middle and
bottom panels of Fig.\ 5 shows the calculated results.  In the figure,
we also perform the same normalization as the ND state of $^{28}$Si to
the experimental results. We consider that our calculated result for the
SD state of $^{32}$S corresponds to that with $N=24$ of Refs.\
\cite{Kim04,Ohk02,koc10}.  After the normalizations, we see that the behaviors
of the calculated results for the SD states of both $^{28}$Si and
$^{32}$S are similar to the experimental moment of inertia proposed by
Refs.\ \cite{Mori85,She82}.  However, further experimental
investigations are necessary for establishing the existence of the SD
states and for band assignment.

At high angular momentum, the asymmetry at the shape configurations of
the ND and SD minima for $^{28}$Si and $^{32}$S become close to the
$^{16}$O+$^{12}$C and $^{16}$O+$^{16}$O divisions, respectively.  In
Figs.~3 and 4, we can clearly see drastic shape transitions, that is,
from densities with one-center to two-center cluster-like configuration.
For the ND and SD minima for $^{28}$Si and $^{32}$S, the neck formation
occurs suddenly at $I=10$ and 8, respectively.

\begin{figure}[t]
\includegraphics[keepaspectratio,width=\linewidth]{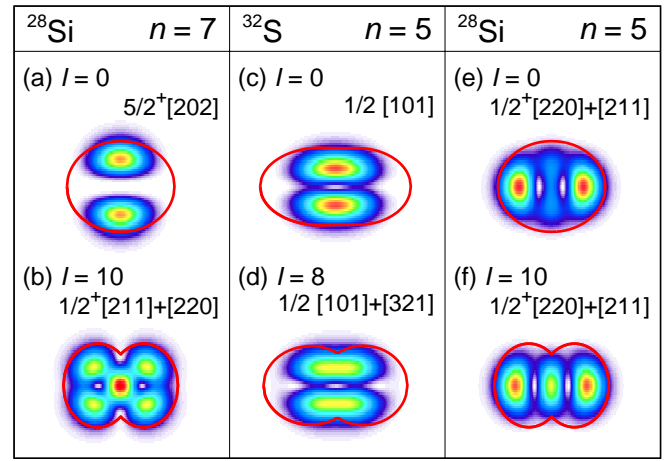}\\%
\caption{(Color online) Density distributions of neutron single-particle wave
 functions for $^{28}$Si and $^{32}$S. The solid line denotes the half
 depth of the mean-field potential. We normalize the color to the
 density distribution at the maximum of each plot at $I=0$.}
\end{figure}

There are two important mechanisms for such clusterization: (i)
intersection between a high-$\Omega$ level, whose energy increases with
deformation and is mainly localized in the ``equator'' region, and a
low-$\Omega$ level whose energy decreases with deformation (intruder
level) and is mainly localized in the ``polar'' regions, (ii) mixing of
single-particle levels with high quantum number.  The former can be seen
in the shape transition of the ND minimum for $^{28}$Si. At $I=0$, the
neutron level at the Fermi surface consists of the $5/2^+[202]$
component ((a) in Fig. 6), which forms the surface of the middle body
part in the total density.  At $I=10$, an transition occurs between this
last occupied level and the intruder level, which is an admixture of
$1/2^+[211]$ (75.3\%) and [220] (16.4\%) and which now becomes the
highest occupied level as shown in (b) in Fig.~6.  In this case, the
wave-function density shifts from the surface of the middle body into
the two nascent fragments.  The second mechanism (ii) is at play in the
SD minimum for $^{32}$S. At $I=0$, the neutron single particle at the
seventh level consists of the 1/2[101] component ((c) in Fig. 6).  At
$I=8$, the components of [321] is slightly mixed in this level.  The
single-particle density of the middle body becomes low due to
1/2$^-$[101] (93.2\%) and [321] (4.3\%), as shown in (d) in Fig.~6.  The
component of [321] describes stretching of the single-particle densities
and [321] is strongly fragmented into many of the levels in both
$^{28}$Si and $^{32}$S.

The other interesting behavior of the single-particle wave function that
influences the clusterization is the neck formation. The neutron
single-particle wave function of the fifth level for $^{28}$Si consists
of the 1/2$^{+}$[220] (66.9\%) and [211] (26.8\%) components ((e) in
Fig. 6). With increasing angular momentum, the [220] component
increases, whereas the [211] component decreases, which forms the neck
part between two fragments. At $I=10$, the wave function is described by
1/2$^+$[220] (80.3\%) and [211] (17\%), as shown in (f) in Fig. 6.  This
trend can be also seen in the [220] component for $^{32}$S.  Although it
seems that two fragments are well separated at the high angular
momentum, those are tightly bonded by the neck formation.

In more elaborate microscopic calculations, the lowest level $J=0^+$
would contain components of intrinsic states with different $I$.  The
highest $J$ is limited to what is obtained when all the spins are
aligned, although the potential pocket still exists at high $I$.  In
this respect, the obtained density distribution at the ND minimum of
$I=10$ for $^{28}$Si is very similar to that at $J=0^{+}$ of
Ref.~\cite{Tani09}.  Consequently, the ND and SD states can contain
cluster components even at $J=0^+$.

\begin{figure}[t]
\includegraphics[keepaspectratio,width=\linewidth]{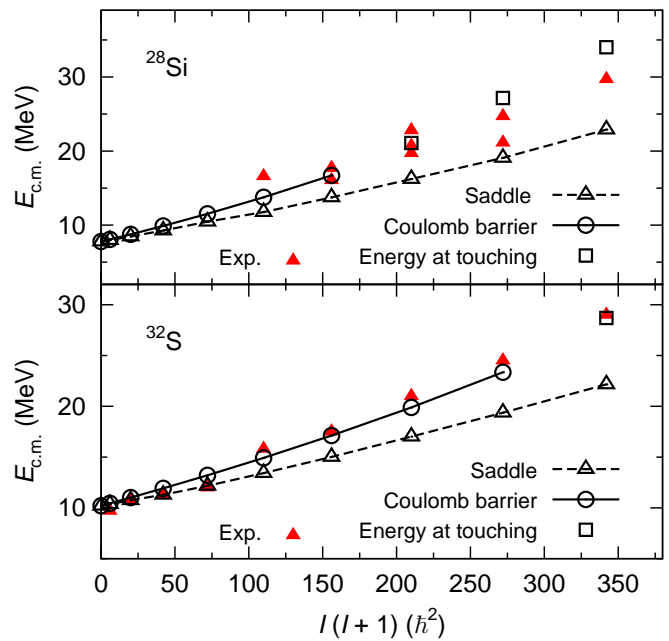}\\%
\caption{Heights of the Coulomb barrier versus the orbital angular
 momentum in the $^{16}$O+$^{12}$C (the top panel) and the
 $^{16}$O+$^{16}$O (the bottom panel) reactions.
 The potential energy is
 measured from the infinitely separated two nuclei.
 The solid line with
 the circles denotes the calculated heights of the Coulomb barrier. The
 open square denotes the energy at the touching point when the potential
 pocket in the fusion barrier vanishes.  The solid triangle denotes the
 average energies of experimental data for the molecular-resonance
 states at each angular momentum taken from
 Refs.~\cite{Ohk02,Ohk04}.  The dashed line with the open
 triangles denotes the heights of the saddle points leading to the
 $^{16}$O+$^{12}$C and $^{16}$O+$^{16}$O reaction channels in the
 calculated potential-energy surface for $^{28}$Si and $^{32}$S,
 respectively. The calculated heights of the saddle points are shifted
 so as to fit to the Coulomb-barrier height at $\ell=0$.}
\end{figure}

Our calculations show a plausible mechanism for the origin of the
molecular resonances. In the calculations, we can identify the potential
valleys leading to the $^{24}$Mg+$\alpha$ and $^{16}$O+$^{12}$C channels
in $^{28}$Si and to $^{28}$Si+$\alpha$ and $^{16}$O+$^{16}$O channels in
$^{32}$S, as shown in Figs.~1 and 2. Also, in this study we cannot
clearly identify any other valleys and associated density clusters in
the potential surface.  Consequently there is an interesting 
correspondence between the valley structures obtained in our
calculations and the observed reaction channels associated with
molecular resonances. Expressed differently we could say that when
entrance-channel target/projectile mass ratios are similar to 
the one-body density clusters corresponding to the calculated valleys in
the potential-energy surfaces do we experimentally observe molecular
resonances.

As shown in this study, the highly excited SD and ND states for
$^{28}$Si and $^{32}$S contain significant $^{16}$O+$^{12}$C and
$^{16}$O+$^{16}$O cluster components, respectively.  We expect that
those two states relate to the observed molecular resonances, because
their mass asymmetry at high angular momentum are very close to the
target/projectile combinations in the entrance channel for which
molecular-resonances are observed.  It is thus interesting to
investigate how those states in the one-body system relate to the
molecular-resonances in the two-body reaction channels.

A key question is whether the molecular resonances arise because of
effects after fusion, during formation of the compound system or because
of effects in the final stages of the two-body heavy-ion collision.  The
molecular resonances emerge just below the Coulomb barrier in the
two-body reaction channels, indicating that those states exist in the
region of slightly overlapping densities of colliding two nuclei.  It is
thus unclear whether those two nuclei are strongly or weakly coupled
with each other, corresponding to the one-body ``sticking'' or the
two-body ``freely rotating'' limits~\cite{Tsa74}, respectively.  To
investigate those two limits, we calculate the Coulomb barrier heights
of the freely rotating and sticking limits, and investigate the
correlation between those and the molecular-resonance states.

For the freely rotating limit, we calculate the Coulomb-barrier heights
as a function of the orbital angular momentum $I$ for the
$^{16}$O+$^{12}$C and $^{16}$O+$^{16}$O reactions.  In the calculation,
we use the Yukawa-plus-exponential model, which is the same framework as
used in the present calculations of the potential-energy surface and is
well tested in many two-body reactions~\cite{Kra79,ichi-fusion}. The
Coulomb interaction energy is calculated for two point charges. The
centrifugal potential is $\hbar^2I(I+1)/2\mu r^2$, where $\mu$ is the
reduced mass and $r$ is the center-of-mass distance between colliding
nuclei. That is, the moment of inertia in the two-body system,
$\mathscr{J}_{(\rm 2bd)}$, is given by $\mathscr{J}_{(\rm 2bd)}=\mu
r^2$, corresponding to the rotational energy of two freely rotating
rigid bodies.  When the fusion barrier does not go over a maximum during
the approach of the two colliding heavy ions, that is it keeps rising
until touching, we follow conventional practice and define the ``Coulomb
barrier'' as the energy at touching.  In the sticking limit, the Coulomb
barrier heights correspond to that of the saddle point in the calculated
one-body potential-energy surface.

Figure 7 shows the resulting Coulomb-barrier heights measured
relative to two infinitely separated nuclei (the open circle with the
solid line). The energy at the touching point is denoted by the open
square.  
The average energies of experimental data at
each $\ell$ for the molecular-resonance states tabulated in Refs.~\cite{Ohk02,Ohk04} are denoted by solid triangles. For
comparison, we also plot the height of the saddle points leading to the
$^{16}$O+$^{12}$C and $^{16}$O+$^{16}$O reaction channels in the
calculated potential-energy surface for $^{28}$Si and $^{32}$S (the open
triangle with the dashed line).  We shift the calculated height of the
saddle points so as to fit to the Coulomb-barrier height of the two-body
reactions at $I=0$, because we here focus on discussing their moments of
inertia, not its absolute energies.

In the figure, we can clearly see the calculated Coulomb-barrier heights
strongly correlate with the experimental data of the molecular-resonance
states, whereas the slope of the height of the saddle points differs
from those. The moment of inertia for the molecular-resonance states is
well reproduced by the freely rotating $\mathscr{J}_{(\rm 2bd)}$, rather
than the one-body ridged rotor $\mathscr{J}_\perp$, indicating that the
molecular resonances are governed by effects in the two-body entrance
channel.

In the deformed states, the two clusters show the property of the
one-body ridged rotor as shown in Fig.~5, whereas in the
molecular-resonance states, they can freely rotate.  The former comes
from the single-particle wave functions tightly bonding two clusters as
shown in Fig.~6 (b) and (f).  That is, the one-body ridged rotor would
change to the two-body freely rotating rotor, if such bonding wave
functions of the neck part vanish by the development of two clearly
separated clusters.  Such wave functions thus play an important role in
transitioning from the one-body deformed and the two-body
molecular-resonance states.

In summary, we have investigated cluster formation in the one-body ND
and SD states for $^{28}$Si and $^{32}$S and their relation to the
molecular resonance states.  We calculated the total-energy surfaces
with inclusion of a rotational-energy contribution in a five-dimensional
deformation space and analyze these as functions of angular momentum.
We identify the ND and SD minima in the potential-energy surface for
$^{28}$Si and $^{32}$S.
The obtained deformed minima are quite similar to those proposed by
other theoretical models.
The level spacings of the rotational bands for
those deformed minima are in good agreement with the experimental data.
The nuclear densities in the ND and SD minima become very cluster-like
when the angular momentum reaches $I=8$ and $10$, respectively.  We show
how cluster configurations develop due to changing occupation of
specific single-particle levels with increasing deformation and angular
momentum.  When we consider the paths from the one-body ND and SD states
to the $^{12}$C+$^{16}$O and $^{16}$O+$^{16}$O channels and change the
inertia from that of 
a one-body rigid rotor to that of a freely rotating system for the
corresponding two-body reaction channels, we can show that the molecular
resonances are connected to the ND and SD states.

\begin{acknowledgments}
 This work was done in the Yukawa International Project for Quark-Hadron
 Sciences (YIPQS), and was partly supported by the GCOE program ``The
 Next Generation of Physics, Spun from Universality and Emergence'' from
 MEXT of Japan.  The work of Y.K-E. was supported by JPSJ Kakenhi
 (No.~22540275).  P.M. acknowledges that this work was carried out under
 the auspices of the National Nuclear Security Administration of the US
 Department of Energy at Los Alamos National Laboratory under Contract
 DE-AC52-06NA25396 and was also supported by a travel grant to JUSTIPEN
 (Japan-US Theory Institute for Physics with Exotic Nuclei) under Grant
 DE-FG02-06ER41407 (U. Tennessee).  The numerical calculations were
 carried out on Altix3700 BX2 at YITP in Kyoto University.
\end{acknowledgments}                     

\end{document}